\documentclass[12pt]{article}
\usepackage{graphicx}
\usepackage{url}
\usepackage{hyperref}
\usepackage{epstopdf}
\usepackage[printonlyused]{acronym}
\usepackage{subfig}
\usepackage{amssymb}
\usepackage{amsmath}
\usepackage{amsfonts}
\DeclareMathOperator*{\argmin}{argmin}

\usepackage[utf8]{inputenc} 
\usepackage{geometry}
\title{Hypergraph based Subnetwork Extraction using Fusion of Task and Rest Functional Connectivity}
\author{Chendi Wang, Rafeef Abugharbieh}  %\inst{1}
\date{}

\begin{document}
% %
% \frontmatter          % for the preliminaries
% %
% %\pagestyle{headings}  % switches on printing of running heads
% \addtocmark{Hypergraph Taskrest} % additional mark in the TOC
% %
% \title{Fusion of Task and Resting State Functional Connectivity for Subnetwork Extraction Based on Hypergraph}
% %
% \titlerunning{Modularity Reinforcement }  % abbreviated title (for running head)
% %                                     also used for the TOC unless
% %                                     \toctitle is used
% %
% \author{Chendi Wang, Rafeef Abugharbieh}  %\inst{1}
% % index{Wang, Chendi}
% % index{Abugharbieh, Rafeef}

% %
% \authorrunning{Chendi Wang et al.} % abbreviated author list (for running head)

% %%%% list of authors for the TOC (use if author list has to be modified)
% %\tocauthor{}
% %
% \institute{Biomedical Signal and Image Computing Lab, UBC, Canada\\
% \email{{chendiw,rafeef}@ece.ubc.ca}}

\maketitle              % typeset the title of the contribution

% This file provides examples of some useful macros for typesetting
% dissertations.  None of the macros defined here are necessary beyond
% for the template documentation, so feel free to change, remove, and add
% your own definitions.
%
% We recommend that you define macros to separate the semantics
% of the things you write from how they are presented.  For example,
% you'll see definitions below for a macro \file{}: by using
% \file{} consistently in the text, we can change how filenames
% are typeset simply by changing the definition of \file{} in
% this file.
% 
%% The following is a directive for TeXShop to indicate the main file
%%!TEX root = diss.tex

\newcommand{\NA}{\textsc{n/a}}	% for "not applicable"
\newcommand{\eg}{e.g.,\ }	% proper form of examples (\eg a, b, c)
\newcommand{\ie}{i.e.,\ }	% proper form for that is (\ie a, b, c)
\newcommand{\etal}{\emph{et al.}}

% Some useful macros for typesetting terms.
\newcommand{\file}[1]{\texttt{#1}}
\newcommand{\class}[1]{\texttt{#1}}
\newcommand{\latexpackage}[1]{\href{http://www.ctan.org/macros/latex/contrib/#1}{\texttt{#1}}}
\newcommand{\latexmiscpackage}[1]{\href{http://www.ctan.org/macros/latex/contrib/misc/#1.sty}{\texttt{#1}}}
\newcommand{\env}[1]{\texttt{#1}}
\newcommand{\BibTeX}{Bib\TeX}

% Define a command \doi{} to typeset a digital object identifier (DOI).
% Note: if the following definition raise an error, then you likely
% have an ancient version of url.sty.  Either find a more recent version
% (3.1 or later work fine) and simply copy it into this directory,  or
% comment out the following two lines and uncomment the third.
\DeclareUrlCommand\DOI{}
\newcommand{\doi}[1]{\href{http://dx.doi.org/#1}{\DOI{doi:#1}}}
%\newcommand{\doi}[1]{\href{http://dx.doi.org/#1}{doi:#1}}

% Useful macro to reference an online document with a hyperlink
% as well with the URL explicitly listed in a footnote
% #1: the URL
% #2: the anchoring text
\newcommand{\webref}[2]{\href{#1}{#2}\footnote{\url{#1}}}

% epigraph is a nice environment for typesetting quotations
\makeatletter
\newenvironment{epigraph}{%
	\begin{flushright}
	\begin{minipage}{\columnwidth-0.75in}
	\begin{flushright}
	\@ifundefined{singlespacing}{}{\singlespacing}%
    }{
	\end{flushright}
	\end{minipage}
	\end{flushright}}
\makeatother

% \FIXME{} is a useful macro for noting things needing to be changed.
% The following definition will also output a warning to the console
\newcommand{\FIXME}[1]{\typeout{**FIXME** #1}\textbf{[FIXME: #1]}}

% END

\setcounter{page}{1}
\pagenumbering{arabic}
\begin{center}
Biomedical Signal and Image Computing Lab, UBC, Canada

chendiw@ece.ubc.ca, rafeef@ece.ubc.ca
\end{center}
\begin{abstract}
Functional subnetwork extraction is commonly used to explore the brain\textquoteright s modular structure. However, reliable subnetwork extraction from functional magnetic resonance imaging (fMRI) data remains challenging due to the pronounced noise in neuroimaging data. In this paper, we proposed a high order relation informed approach based on hypergraph to combine the information from multi-task data and resting state data to improve subnetwork extraction. Our assumption is that task data can be beneficial for the subnetwork extraction process, since the repeatedly activated nodes involved in diverse tasks might be the canonical network components which comprise pre-existing repertoires of resting state subnetworks \cite{Park-2014-Graph}. Our proposed high order relation informed subnetwork extraction based on a strength information embedded hypergraph, (1) facilitates the multisource integration for subnetwork extraction, (2) utilizes information on relationships and changes between the nodes across different tasks, and (3) enables the study on higher order relations among brain network nodes.
On real data, we demonstrated that fusing task activation, task-induced connectivity and resting state functional connectivity based on hypergraphs improves subnetwork extraction compared to employing a single source from either rest or task data in terms of subnetwork modularity measure, inter-subject reproducibility, along with more biologically meaningful subnetwork assignments.

\end{abstract}
\section{Introduction}
The human brain can be regarded as being a network where units, or nodes, represent different specialized regions, and edges represent communication pathways. Brain network analysis methods for connectome studies include an important branch of brain subnetwork identification. Given brain connectivity matrices, brain networks can be quantitatively examined for certain commonly used network measures. The modular structure (community structure) is of particular interest; it is from this structure that we can infer information about brain subnetworks. The modular structure is extracted by subdividing a network into groups of nodes with the maximal possible within-group links and minimal between-group links using community detection methods \cite{Girvan-2002-Community}.

Most existing functional subnetwork extraction methods focus on resting state function connectivity data \cite{Van-2008-Normalized,Nicolini-2016-Modular}, using functional homogeneity clustering, \ac{ICA}, or graph community detection. However, resting state functional connectivity is inherently with low \ac{SNR} and prone to false positive correlations \cite{Murphy-2013-fMRI}. Such noisy resting state functional connectivity information leads to unreliable subnetwork extraction results. Given the resemblance between resting state and task functional subnetworks \cite{Smith-2009-Correspondence} and high order nodal relations reflected from multi-task data, we here aim to incorporate information from task data into the subnetwork extraction based on multilayer network. We explore if this integration can improve the subnetwork extraction by exploiting the mechanism of how groups of nodes collaborate together to execute a function and how these groups communicate with each other.
\subsection{Related Work - Relationship between Task and Resting Functional Connectivity}
\label{subsec:task_re}
Recent studies indicate that resting state functional activity actually persists during task performance \cite{Fox-2007-Intrinsic}, and similar network architecture is present across task and rest, which is supported by the existence of similar multi-task \ac{FC} and resting-state \ac{FC} matrices that were averaged across subjects 
% (with correlation R = 0.90, $p<0.00001$ based on 40 minutes of task \ac{fMRI} data for 7 tasks and 56 minutes of resting \ac{fMRI} data per subject from 118 \ac{HCP} subjects.) 
\cite{Cole-2014-Intrinsic}. % with different preprocessing, we find it to be R(C(taskAll7, REST))= 0.7845; p<0.05; across subject...77 subject
% A recent study identified a whole-brain resting-state network architecture that was similar to the network architecture present across dozens of task states \cite{Cole-2014-Intrinsic}.
Studies have also shown that there is a strong resemblance between rest and task subnetworks \cite{Smith-2009-Correspondence,Sporns-2016-Modular}. The spatial overlap between resting-state functional subnetworks and task-evoked activities has been discovered \cite{Tavor-2016-Task,Chan-2017-Resting}.
 
Based on the close relationship between the two, resting state data have been used to predict the task activities, by using group \ac{ICA} to discover repertories of canonical network components that will be recruited in tasks \cite{Park-2014-Graph}; by applying the graphical connectional topology of brain regions at rest to predict functional activity of them during task \cite{Chan-2017-Resting}; or based on a voxel-matched regression method to estimate the magnitude of task-induced activity \cite{Mennes-2010-Inter}. 

On the other hand, aggregating brain imaging data from thousands of task related studies allowed the construction of ‘co-activation networks’, whose major components and overall network topology strongly resembled functional subnetworks derived from resting-state recordings \cite{Crossley-2013-Cognitive,Bertolero-2015-Modular,Bassett-2017-Network}.

It has been suggested that networks involved in cognition are a subset of networks embedded in spontaneous activity \cite{Smith-2009-Correspondence,Laird-2011-Behavioral}, and a number of canonical network components in the pre-existing repertoires of intrinsic subnetworks are selectively and dynamically recruited for various cognitions \cite{Mesulam-1998-Sensation,Park-2014-Graph}. 
%%%%%%\cite{Park-2014-Graph} overlapping method, need to mention or compare with our method.
%%%%%% an edge can be within different subnetworks with different weights – similar to hypergraph idea? Need to Compare??
% Page 18 “confirming earlier work on the similarity of task- evoked and resting brain networks (Smith et al. 2009).” [19]
% Page 3 “The correspondence between coactivation and connectivity network modular decompositions was high, with a Rand index of 0.78 (significantly greater than the correspondence between modules of the connectivity net- work and randomly reassigned modules of the coactivation network” [20]
%[21] Brainmap Database and resting state Fmri: “these cognitive components have a spatial distribution that is qualitatively similar to the modules identified from a network analysis of spontaneous neural activity.”
% [2] the across-subject mean resting-state FC and multi-task FC matrices were highly similar (r=0.90, p<0.00001), supporting the existence of intrinsic FC common across rest and a variety of task states.
% a highly similar underlying network architecture is present across rest and task.
\subsection{Related Work - Multilayer Brain Network Analysis}
Multilayer network has recently been used to model and analyze complex high order data, such as multivariate and multiscale information within the human brain \cite{De-2017-Multilayer}. Different layers can represent relationships across different temporal variations \cite{Muldoon-2016-Network}, reflect different imaging modalities (such as task and rest) \cite{De-2017-Multilayer}, or different frequency bands \cite{De-2016-Mapping}, etc. Hypergraph is a type of multilayer graphs, in which edges can link any number of nodes \cite{Zhou-2007-Learning}. Hypergraphs have been used to identify non-random structure in structural connectivity of the cortical microcircuits \cite{Dotko-2016-Topological}, identify high order brain connectome biomarkers for disease prediction \cite{Zu-2016-Identifying}, and study relationships between functional and structural connectome data \cite{Munsell-2016-Identifying}.
\section{High Order Relation Informed Subnetwork Extraction}
\label{subsec:Task_hyper}
% To improve the subnetwork extraction based on resting state functional data alone, which lacks high order relation information, we propose to incorporate task data into the process based on hypergraph.
% Considering the significant correspondence between resting and task functional subnetwork patterns, we propose to use hypergraph to perform subnetwork extraction by fusing task data into the resting state data based approach,  
Our assumption is that task data can be beneficial for subnetwork extraction since the repeatedly activated nodes in different tasks could be the canonical network components in the spontaneous resting state subnetworks. At the same time, the multilayer structure of repeatedly activated nodes across multi-task can be elegantly presented as a hypergraph. We propose a high order relation informed subnetwork extraction model, which (1) facilitates multisource integration of task and rest data for subnetwork extraction, (2) utilizes information from the relationship between groups of activated nodes across different tasks, and (3) enables the study on higher order relations among brain network nodes.
% ((4) whole brain coverage from resting state \ac{FC}?)
\subsection{Framework}
We propose a high order relation informed approach based on hypergraph to integrate both resting state and task information for brain subnetwork extraction. We firstly construct a brain graph based on a certain parcellation atlas. Secondly, we detect activation of brain nodes from task data to define the nodes for multiple layers in the hypergraph, and define the connection strength between nodes using task-induced connectivity. Thirdly, we construct the multitask hypergraph and incorporate resting state \ac{FC} strength information when setting the weights of hyperedges. Fourthly, we fuse task and rest \ac{FC} using weighted combination model % an automatic adaptive weighting term ($\gamma$ based on the number of tasks), - save for future work.
before performing graphcut on the constructed graph.
% , or a multislice community detection method \cite{Mucha-2010-Community} .
%
%Add a figure for the framework here.
%
\subsection{Notation Overview of Hypergraph} %  and Graphcut on Hypergraph
\subsubsection{Notations}
\label{para:Nota_hyper}
\begin{figure}
	\centering
	\subfloat[Toy example of a hypergraph]{
        \includegraphics[height=1.4in]{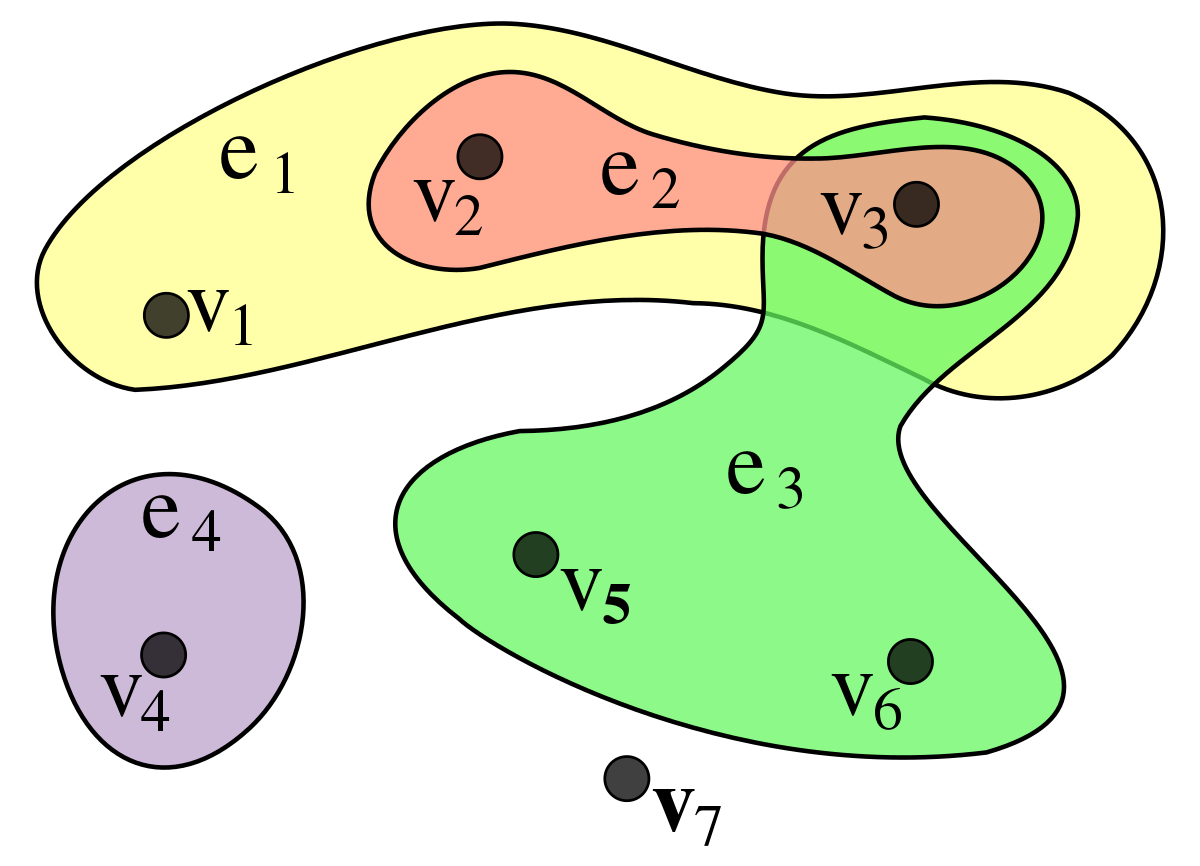}
    }
    \subfloat[Simple graph]{
        \includegraphics[height=1.4in]{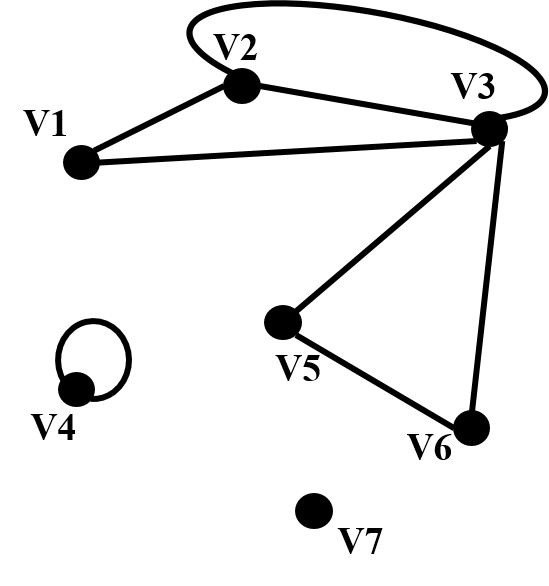}
    }
    \subfloat[Incidence matrix H]{
        \includegraphics[height=1.4in]{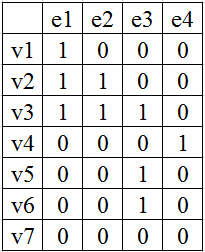}
    }
    \caption{Hypergraph and its corresponding simple graph and incidence matrix. Left: an hyperedge set $E = \{e_1, e_2, e_3, e_4\}$ and a node set $V = \{v_1, v_2, v_3, v_4, v_5, v_6, v_7\}$. Middle: the corresponding simple graph. Right: the incidence matrix $\mathbf{H}$ of the hypergraph on the left, with the entry $(v_i,e_j)$ being set to 1 if $v_i$ is in $e_j$, and 0 otherwise.}
    % We can see that nodes $v_4$ and $v_7$ are presented quite different in the hypergraph, but there is no difference when they are presented in a simple graph.
    % add the difference between v4 and v7. simple graph can't tell from not activated node v7, and activated node v4.
    \label{fig:Hypertoy}
\end{figure}

We here follow most of the notations presented in \cite{Zhou-2007-Learning}. Let $V$ denote a set of nodes, and $E$ denote a family of subsets $e$ of $V$ such that $\cup e\in E = V$. Then we define $G = (V;E)$ a hypergraph with the vertex set $V$ and the hyperedge set $E$. A hyperedge containing just two nodes is a simple graph edge. A hyperedge $e$ is said to be incident with a node $v$ when $v \in e$. Two nodes are \textit{connected} if they both belong to the same hyperedge. Two hyperedges are connected if the intersection of them is not an empty set, $e_i\cap e_j \neq \emptyset$. Given an arbitrary set $X$, let $|X|$ denote the cardinality of $X$.
A hypergraph $G$ can be represented by a $|V|\times |E|$ \textit{incidence} matrix $\mathbf{H}$ with entries $h(v, e) = 1$ if $v \in e$ and 0 otherwise, see an example in \autoref{fig:Hypertoy}.
A weighted hypergraph, $G = (V;E;w)$, is a hypergraph that has a positive number $w(e)$ associated with each hyperedge $e$, called the weight of hyperedge $e$. Next, we define four important measures of hypergraph properties.

For a hyperedge $e\in E$:

1. We follow \cite{Zhou-2007-Learning} to define its degree as $d\_h(e)=\delta(e):=|e|$, which counts the number of nodes that exist in the hyperedge. If one uses the incidence matrix, $\delta(e): = \sum_{\{v\in V\}}h(v,e)$. Let $\mathbf{D_e}$ denote the diagonal matrices containing the hyperedge degrees. Take \autoref{fig:Hypertoy} as an example, $\delta(e_1) = 3$, and $\delta(e_2) = 2$.

2. We further define the hyperdegree of a hyperedge as the number of hyperedges connected to it, denoted as $d\_hH(e): = \sum_{\{e_i\in E, e_i\neq e\}} e\cap e_i$. For example, $d\_hH(e_1) = 3$, $d\_hH(e_3) = 2$, and $d\_hH(e_4) = 0$ in \autoref{fig:Hypertoy}.  

For a node $v \in V$:

% 3. We follow \cite{Berge-1984-Hypergraphs} to define its degree by $d(v): = \sum_{\{e\in E\}}h(v,e)$, which counts the number of hyperedges which include this node. Let $\mathbf{D_v}$ denote the diagonal matrices containing the node degrees. For example, $d(v_3) = 3$ and $d(v_7) = 0$ when take the graph as binary in \autoref{fig:Hypertoy}.
3. We follow \cite{Zhou-2007-Learning} to define its degree by $d(v) = \sum_{\{e\in E | v \in e\}} w(e)$. If one uses the incidence matrix, $d(v) = \sum_{\{e\in E\}}w(e)h(v,e)$. When all $w(e)$ = 1, $d(v)$ counts the number of hyperedges which include this node: $d(v) = \sum_{\{e\in E | v \in e\}} 1$, or $d(v) = \sum_{\{e\in E\}}h(v,e)$. Let $\mathbf{D_v}$ denote the diagonal matrices containing the node degrees.

4. We then define the hyperdegree of a node as $d\_H(v): = \sum_{\{v\in e | e\in E\}} \delta(e)$, which counts the number of nodes connected to a particular node across all hyperedges. For example, $d\_H(v_2) = 5$, $d\_H(v_3) = 6$, $d\_H(v_5) = 3$ in \autoref{fig:Hypertoy}. Its weighted version will be estimating the strength between the connected node pairs. 
% switch hyperdegree and degree

Next, let $\mathbf{W}$ denote the diagonal matrix containing the weights $w(e)$ of hyperedges. Correspondingly, the adjacency matrix $\mathbf{A}$ of hypergraph $G$ is defined as:
\begin{equation}
\label{eq:hyper_A}
\mathbf{A} = \mathbf{HWH}^{T} - \mathbf{D_v}, 
\end{equation}
where $\mathbf{H}^{T}$ is the transpose of $\mathbf{H}$.
\subsubsection{Graphcut of the Hypergraph}
One can group the nodes into subsets using graph partitioning methods, \ie graphcut. The intuition is to find a partition of the graph such that the edges within a subset have high weights (strong intra-class connections), and the edges between different subsets have low weights (weak inter-class connections). Let $S \in V$ denote a subset of nodes and $S^c$ denote the complement of $S$. Follow the notations in \cite{Von-2007-Tutorial}, the adjacency matrix $\mathbf{A}(X,Y):=\sum_{i\in X, j\in Y} a_{ij}$. For a given number $M$ of subnets, the Mincut approach \cite{Stoer-1997-Simple} implements the graphcut by generating a partition $S_1,\ldots,S_M$ which minimizes
\begin{equation}
\label{eq:Mincut}
\text{cut}(S_1,\ldots,S_M) :=\frac{1}{2} \sum^M_{i=1}\mathbf{A}(S_i,S^c_i).
\end{equation}

To solve the problem of separating individual nodes as a subset in Mincut, RatioCut \cite{Hagen-1992-New} and \ac{Ncuts} \cite{Shi-2000-Normalized} have been proposed to encode the information of the size of a subset.
\begin{equation}
\label{eq:Ratiocut}
\text{RatioCut}(S_1,\ldots,S_M) :=\frac{1}{2} \sum^M_{i=1}\frac{\mathbf{A}(S_i,S^c_i)}{|S_i|} = \sum^M_{i=1}\frac{\text{cut}(S_i,S^c_i)}{|S_i|},
\end{equation}
where $|S|$ measures the number of nodes in $S$.
\begin{equation}
\label{eq:Ncuts1}
\text{Ncut}(S_1,\ldots,S_M) :=\frac{1}{2} \sum^M_{i=1}\frac{\mathbf{A}(S_i,S^c_i)}{\text{vol}(S_i)} = \sum^M_{i=1}\frac{\text{cut}(S_i,S^c_i)}{\text{vol}(S_i)},
\end{equation}
where $\text{vol}(S)$ measures the volume of $S$ by summing over the weights of all edges attached to the nodes as $\text{vol}(S) :=\sum_{v\in S}ds(v)$, and node strength $ds(v)$ is the weighted version of node degree $d(v)$.

\ac{Ncuts} has been widely used in image segmentation and brain study community, since it utilizes the weight information. In the following, we show that \ac{Ncuts} approach can be generalized from simple graphs to hypergraphs, which has been proven in \cite{Zhou-2007-Learning}.

For a hypergraph $G = (V;E;w)$, a cut is a partition of $V$ into two parts $S$ and $S^c$. A hypergraph $e$ is cut when it is incident with the nodes in $S$ and $S^c$ at the same time. The \textit{hyperedge boundary} of $S$ is defined as $\partial S:=\{e\in E|e\cap S \neq \emptyset, e\in E|e\cap S^c \neq \emptyset\}$, which is a hyperedge set consisting of the hyperedges which are cut \cite{Zhou-2007-Learning}. The definition of the volume in a hypergraph $\text{vol}(S)$ is the sum of the degrees of the nodes in $S$, $\text{vol}(S) :=\sum_{v\in S}d(v)$. Each hyperedge is essentially a fully connected subgraph, then the edges in a subgraph is called subedges, being assigned with the same weight $w(e)/\delta(e)$. When a hyperdege $e$ is cut, there are $|e\cap S||e\cap S^c|$ subedges are cut. Hence, the volume of $\partial S$ is defined by 
\begin{equation}
\label{eq:hyperbound}
\text{vol}(\partial S) :=\sum_{e\in \partial S}w(e)\frac{|e\cap S||e\cap S^c|}{\delta(e)},
\end{equation}
which is the sum of weights over the subedges being cut.
By this definition, we have $\text{vol}(\partial S) = \text{vol}(\partial S^c)$. Similar to the simple graphs, Normalized hypergraph cut is to keep the high intra-class connection and low inter-class connection with a partition $S_1,\ldots,S_M$ by minimizing the cut as below:
\begin{equation}
\label{eq:hypercut}
\argmin_{\emptyset \neq S_1,\ldots,S_M\subset V} \sum_{i=1}^M \frac{\text{vol}(\partial S_i)}{\text{vol}(S_i)}.
\end{equation}
\subsection{Task Activation Detection - Node Definition in the Hypergraph}
\label{para:task_activation}
In order to construct the multiple layers in the hypergraph, we apply the activation detection technique on the task data to define the nodes that are contained in different hyperedges. 
% Then, each hyperedge will be a fully connected subgraph comprising of the activated nodes in each task.
The standard way of activation detection is to use a \ac{GLM} where statistics, such as t-values, reflect the degree of the similarity between the stimulus and voxel time courses. The estimated statistics produce an activation statistics map (t-map), followed by a thresholding of the map to identify the activated voxels \cite{Friston-1994-Statistical}. Due to the pronounced noise in the \ac{fMRI} data, activation detection at the individual level could be inaccurate \cite{Ng-2012-Modeling}. In order to derive more reliable task-induced activation, we have chosen a group activation detection over the individual based approach. First, to compute the intra-subject activation patters, a standard \ac{GLM} is applied as below \cite{Friston-1994-Statistical}:
\begin{equation}
\label{eq:activation}
\mathbf{Y}^i = \mathbf{X}^i \beta^i + \mathbf{E}^i,
\end{equation}
where $\mathbf{Y}^i$ is a $t\times N$ matrix of the task-induced \ac{fMRI} time courses of $N$ brain regions from subject $i$, $\beta^i$ is a $d\times N$ activation matrix to be estimated, $\mathbf{E}^i$ is a $t\times N$ residual matrix, and $\mathbf{X}^i = [\mathbf{X}_{\text{task}}|\mathbf{X}^i_{\text{confounds}}]$ is a $t\times d$ matrix. $\mathbf{X}_{\text{task}}$ is the task regressors and $\mathbf{X}^i_{\text{confounds}}$ is the confound regressors. Next, we combine the activation results across subjects to assemble a group activation map, which is used to define nodes for each layer of the hypergraph. Specifically, we apply a max-t permutation test \cite{Nichols-2003-Controlling} on $\beta^i$ aggregated from all the subjects, which implicitly accounts for multiple comparisons and control over false detections \cite{Yoldemir-2016-Multimodal}. Group activation is declared at a p-value threshold of 0.05.
\subsection{Strength Informed Weighted Multi-task Hypergraph}
In the beginning of \autoref{subsec:Task_hyper}, we argued that multi-task information can be presented as a hypergraph, with the hyperedges being different tasks, and the nodes in each hyperedge being the brain regions activated in a certain task. In the traditional definition of hypergraph, nodes are connected to each other binarily, \ie the edge weights between a node pair are 1 if they are connected, or 0 otherwise. We here propose a strength informed weighted hypergraph model by incorporating the strength information from the connections between nodes. We further determine the hyperedge weight $w(e)$ using the graphical measures defined in \autoref{para:Nota_hyper}.
\subsubsection{Pairwise Nodal Connection Strength Estimation}
\label{para:C_task_comp}
In order to estimate the strength of the connections between two nodes, we use the Pearson's correlations between time courses from pairs of brain regions. We denote the resting state connectivity matrix as $\mathbf{C}^{\text{rest}}$. To produce the task-induced connectivity matrix $\mathbf{C}^{\text{task}}$, we use the task-induced time course information. We follow the strategy in \cite{Cole-2014-Intrinsic} to remove all inter-block rest periods from all regions’ time courses, before computing the pairwise Pearson's correlations across all concatenated block/event duration time courses within a task. To keep the consistency when combining information from the nodes across different layers, we keep all the $\mathbf{C}^{\text{task}}$ having the same dimension of $N\times N$ as the $\mathbf{C}^{\text{rest}}$, then set the rows and columns of non-activated nodes to zero.

\subsubsection{Proposed Strength Informed Weighted Hypergraph}
We present a modified hypergraph cut criteria formulation based on \autoref{eq:hyperbound} to incorporate pairwise nodal connection strength information from $\mathbf{C}$ as below in \autoref{eq:hyperbound_w}. The symbol $\tilde .$ indicates the usage of strength information.
% %
% \begin{equation}
% \label{eq:hypercut_proposed}
% % \argmin_{\emptyset \neq S_1,\ldots,S_M\subset V} \sum_{i=1}^M \frac{\tilde{\text{vol}}(\partial S_i)}{\tilde{\text{vol}}(S_i)},
% \argmin_{\emptyset \neq S_1,\ldots,S_M\subset V} \sum_{i=1}^M \frac{\tilde{\text{vol}}(\partial S_i)}{\text{vol}(S_i)},
% \end{equation}
% %
% % where $\tilde{\text{vol}}(S)$ is the weighted version of $\text{vol}(S)$, defined as $\tilde{\text{vol}}(S) :=\sum_{v\in S}ds(v)$ and $ds(v)$ being the strength of a node derived from each of the task connectivity correlation matrix (each $e$ corresponds to a different task):
% % \begin{equation}
% % \label{eq:node_strength_m}
% % ds(v): = \sum_{\{e\in E | v \in e\}} \sum_{u\in e} \mathbf{C}_{uv}. 
% % \end{equation}
% where $\tilde{\text{vol}}(\partial S)$ is a weighted version of $\text{vol}(\partial S)$ in \autoref{eq:hyperbound},
% %
\begin{equation}
\label{eq:hyperbound_w}
\tilde{\text{vol}}(\partial S) :=\sum_{e\in \partial S}\tilde w(e)\frac{\sum_{i\in \{e\cap S\}, j\in \{e\cap S^c\}} \mathbf{C}^e_{ij}}{\delta(e)},
% = \sum_{e\in \partial S}\frac{w(e)_{c}}{\delta(e)}\sum_{i\in \{e\cap S\}, j\in \{e\cap S^c\}} \mathbf{C}_{ij},
\end{equation}
where $\tilde{\text{vol}}(\partial S)$ is a strength informed version of $\text{vol}(\partial S)$ in \autoref{eq:hyperbound}, $\mathbf{C}^e$ is the connectivity matrix derived from the task corresponding to the layer $e$, and $\tilde w(e)$ is the modified weight item in the hypergraph. We propose here to incorporate strength information from the connectivity matrix and utilize the four hypergraph measures defined in \autoref{para:Nota_hyper} to determine $\tilde w(e)$, whose nature is the importance of the hyperedge in the hypergraph. 
% We hypothesize that the importance of a hyperedge should correspond to the appearance of certain critical roles in a graph, such as hubs.
% and core nodes in overlapping subnetworks. 
Based on the definition of the four hypergraph measures, we exploit their corresponding biological meanings to set $\tilde{\text{vol}}(\partial S)$ and $\tilde w(e)$ as below:

1. The degree of a hyperedge $\delta(e)$ counts the number of brain regions that are activated in a task. To avoid the bias of the hyperedge size, $\tilde{\text{vol}}(\partial S)$ should be normalized by $\delta(e)$.
% Similarly to the simple graphcut, in order to avoid the problem of separating individual nodes as a subset, 
% here we extend $d\_h(e)$ to its weighted version by taking the average value of all the pairwise correlation between a couple of nodes that are in the hyperedge.

2. The hyperdegree of a hyperedge is defined as the number of hyperedges that are connected to it. Higher value indicates that more frequently activated patterns in the brain activities exist in this hyperedge. Thus, $\tilde w(e)$ should be proportional to $d\_hH(e)$, \ie $\tilde w(e) \propto d\_hH(e)$.  

3. The degree of a node counts the number of hyperedges that contain this node, and the biological equivalence is the number of different tasks in which one node is activated. A node with a higher degree is similar to the definition of the connector hubs residing within different subnetworks. Hence, $\tilde w(e)$ should be proportional to some statistics derived from $d(v)$ of the nodes in a hyperedge $e$. We denote the statistics computation method as \textit{stat} here and it can be widely used statistics such as average value (mean), median value (median) and maximum value (max). Thus, $\tilde w(e) \propto stat(d(v))$.

4. The hyperdegree of a node reflects the number of all other nodes that are connected to it across all layers, which equals the number of connections from other co-activated nodes to it across multiple tasks. The biological meaning of a node with a high value coincides with the definition of hubs. Hence, $\tilde w(e)$ should be proportional to some statistics derived from $d\_ H(v)$ of the nodes in a hyperedge $e$, \ie $\tilde w(e) \propto stat(d\_ H(v))$. Here, in order to incorporate strength information, we apply the weighted version of $d\_ H(v)$, the strength of the node $d\_ Hs(v)$ as defined in \autoref{eq:node_strength_m}, \ie $\tilde w(e) \propto stat(d\_ Hs(v))$.
\begin{equation}
\label{eq:node_strength_m}
%d\_ Hs(v): = \sum_{\{e\in E | v \in e\}} \sum_{u, u\in e} \mathbf{C}_{uv}.
d\_ Hs(v): = \sum_{\{v\in e | e \in E\}} \sum_{u\in e} \mathbf{C}^{e}_{uv},
% d\_H(v): = \sum_{\{v\in e\ | e\in E\}} \delta(e)
\end{equation}
where $\mathbf{C}^{e}$ is the task-induced connectivity matrix for the $e$th task.

In order to utilize strength information and hypergraph measures, we propose the $\tilde w(e)$ formulation as below:
\begin{equation}
\label{eq:We_strength}
\tilde w(e): = w1\cdot d\_hH(e) + w2\cdot stat(d(v)) + w3\cdot stat(d\_ Hs(v)),
\end{equation}
where $w1, w2, w3$ are free parameters to control the contributions of each measure to the hyperedge.
% The free parameters and \textit{stat} methods can be optimized by cross validation.

\subsection{Multisource Integration of Rest and Task \ac{fMRI}}
% For the reason!!!.
\label{sec:multi_hyper_m}
Given the close correspondence between task and rest connectivity architecture and subnetworks, we further extend the multi-task hypergraph model to integrate \ac{rs-fcMRI} information. To do that, 
% We have constructed a multi-task hypergraph for subnetwork extraction based on the assumption that repeatedly activated nodes in different tasks could be the canonical network components which comprise pre-existing repertoires of resting state subnetworks in the previous section. We next integrate the resting state \ac{FC} information into the multi-task hypergraph 
we use $\mathbf{C}^{\text{rest}}$ for the pairwise nodal connection strength computation in \autoref{eq:node_strength_m} as below:
% , which is further used to compute in \autoref{eq:We_strength}.
\begin{equation}
\label{eq:node_strength_m_rest}
%d\_ Hs(v): = \sum_{\{e\in E | v \in e\}} \sum_{u, u\in e} \mathbf{C}_{uv}.
d\_ Hs(v): = \sum_{\{v\in e | e \in E\}} \sum_{u\in e} \mathbf{C}^{\text{rest}}_{uv},
% d\_H(v): = \sum_{\{v\in e\ | e\in E\}} \delta(e)
\end{equation}

% !!!Adaptive $\gamma$ depending on the coverage of the tasks!!!.
Furthermore, we explicitly combine the two sources of task and rest data for subnetwork extraction. We firstly fuse the multiple layers of the multi-task hypergraph into one single layer, and secondly combine it with a resting state connectivity layer. 
Given that the hypergraph cut criterion (\autoref{eq:hyperbound}) is to evaluate the aggregated sum of the cuts across all the pairwise subedges (nodal connections) in the hypergragh, 
% and there are multiple edges connecting two nodes that exist across different task layers,
we propose to aggregate the strength information between node pairs across all the layers. To do that, we transform the multiple pairwise nodal connections across task layers (\autoref{eq:hyperbound_w}) into one single nodal connection as below: 
% After we defined the cut criteria defined in \autoref{eq:hypercut} for the multilayer hypergraph, we implement the fusion of the multiple layers of task connectivity matrix modified based on the criteria by taking the average of the $\mathbf{C}$ matrix across all task layers. 
\begin{equation}
\label{eq:multitask_hyper}
\bar{\mathbf{C}}^{\text{task}}_{ij} = \frac{1}{T}\sum_{k = 1}^{T} \frac{\tilde w(e^k)}{\delta e^k}\mathbf{C}_{ij}^{e^k},
\end{equation}
where the subscript $k = 1,\ldots,T$ is the indicator for tasks, $T$ is the total number of tasks available, and $e^k$ is the hyperedge in the $k$th layer of the hypergraph.
% , defined by the activated brain regions in the task.
% corresponds to label $i$. 
$\mathbf{C}^{e^k}$ is the connectivity matrix derived using the time courses in the task $k$ using the procedure described in \autoref{para:C_task_comp}.
% We then apply graphcut approach such as \ac{Ncuts} to this connectivity matrix carrying the information from multi-task.

We next explicitly combine the two sources by a linear weighted combination between the aggregated multi-task connectivity matrix from above (\autoref{eq:multitask_hyper}) and the resting state connectivity matrix in \autoref{eq:multimodal_hyper} as below: 
\begin{equation}
\label{eq:multimodal_hyper}
\mathbf{C}^{\text{t-r}}: = \gamma \bar{\mathbf{C}}^{\text{task}} + (1-\gamma) \mathbf{C}^{\text{rest}},
\end{equation}
where $\gamma$ a free parameter, which can be optimized by cross-validation, or determined by the number of the tasks available. Our linear model for combining two sources, which are both derived from functional modality, was motivated by the study indicating a largely linear superposition of task-evoked signal and resting state modulations in the brain \cite{Fox-2007-Intrinsic}. We also explore combining the two by applying a multislice community detection approach \cite{Mucha-2010-Community}, which extends modularity quality function based on the stability of communities under Laplacian dynamics with a coupling parameter $\omega$ to control over interslice correspondence of communities.
% Based on equivalence between the modularity quality function and stability of communities under Laplacian dynamics, we extended quality function by deriving null models to include multilayer connectivities linked by a coupling parameter. Reweighting of the conditional probabilities, which allows a different resolution gamma.
% Subtracting this conditional joint probability (of motion between two slices) from the linear (in time) approximation of the exponential describing the Laplacian dynamics, we obtained a multislice generalization of modularity.
%
\section{Results}
% Follow the ideas [Cole, 2014];
% We propose to identify brain subnetwork structure using task functional connectivity and activation patterns based on a hypergraph model, which also facilitates integration with resting state functional connectivity.
% Recent studies have demonstrated that the frequent functional connectivity strengths across tasks closely matched the strengths observed at rest, which suggests that there is an “intrinsic” architecture of functional brain organization \cite{Cole-2014-Intrinsic}. Based on this hypothesis, we argue that it is possible to identify this intrinsic and standard subnetwork structure using task data. Since the studies on subnetwork extraction using resting state \ac{FC} is well-established, we here try to first compare the subnetwork results derived using new methods with using only resting state \ac{FC} data, which enables us to see how similar the subnetworks would be and explore new directions to study some new findings. 
We first investigated the similarity of connectivity between resting state  and task-general and task-specific connectivity.  
To evaluate our proposed approaches, we assessed the graphical metric modularity $Q$ value, the inter-subject reproducibility and examined the biological meaning of subnetwork assignments. We applied subnetwork extraction on (1) resting state \ac{FC} alone, (2) task-induced \ac{FC} alone, (3) multi-task hypergraph, (4) multi-task hypergraph integrated with resting state connectivity strength, (5) weighted combination of (4) and resting state \ac{FC}, (6) combination of (4) and resting state \ac{FC} using multislice community detection method \cite{Mucha-2010-Community}.
% 1. Rest C alone
% 2. Task C alone
% 3. Multitask hyper alone
% 4. task+rest hyper alone
% 5. 4+ rest as another layer, weighted combination
% 6. 4 and rest layer using multislice
\subsection{Materials}
We used the resting state \ac{fMRI} and task \ac{fMRI} scans of 77 unrelated healthy subjects from the \ac{HCP} dataset \cite{van-2013-HCP}. Two sessions of resting state \ac{fMRI} with 30 minutes for each session, and 7 sessions of task \ac{fMRI} data were available for multisource integration. The seven tasks are working memory (total time: 10:02), gambling (6:24), motor (7:08), language (7:54), social cognition (6:54), relational processing (5:52) and emotion processing (4:32). 
Preprocessing already applied to the \ac{HCP} \ac{fMRI} data includes gradient distortion correction, motion correction, spatial normalization to \ac{MNI} space with nonlinear registration based on a single spline interpolation, and intensity normalization \cite{Glasser-2013-Minimal}. 
% It was suggested in the \ac{HCP} preprocessing paper \cite{Glasser-2013-Minimal} that no slice timing correction needs to be employed, since the fast \ac{TR} sampling reduces the need for slice timing correction as all slices in each volume are acquired much closer together than in typical \ac{fMRI} acquisitions (\ac{TR} $\sim$ 2.5s).
Additionally, we regressed out motion artifacts, mean white matter and cerebrospinal fluid confounds, and principal components of high variance voxels using compCor \cite{Behzadi-2007-Component}. Next, we applied a bandpass filter with cutoff frequencies of 0.01 and 0.1 Hz for resting state \ac{fMRI} data. For task \ac{fMRI} data, we performed similar temporal processing, except a high-pass filter at 1/128 Hz was used. The data were further demeaned and normalized by the standard deviation.
We then used the \ac{HO} atlas \cite{Desikan-2006-Automated}, which has 112 \ac{ROI}s, to define the brain region nodes. We chose the well-established \ac{HO} atlas because it sampled from every major brain system, and consists of the highest number of subjects with both manual and automatic labelling technique compared to other commonly used anatomical atlases. Voxel time courses within \ac{ROI}s were averaged to generate region time courses. The region time courses were demeaned, normalized by the standard deviation. Group level time courses were generated by concatenating the time courses across subjects. The Pearson's correlation values between the region time courses were taken as estimates of \ac{FC} matrices. Negative elements in all connectivity matrices were set to zero due to the currently unclear interpretation of negative connectivity \cite{Skudlarski-2008-Measuring}.
For task activation, we applied the activation detection on the seven tasks available following the steps described in \autoref{para:task_activation}. 

% we used a limited set of 264 regions of interest to estimate FC throughout the brain (Power et al., 2011). 
% We used this set of regions because it sampled from every major brain system, was estimated using independent data (reducing potential circularity or biases from over-fitting in our analyses (Kriegeskorte et al., 2009)), and came with an independently identified node community partition (again, to reduce potential biases). 
We summarize here the annotation of the graphs for six methods being evaluated for subnetwork extraction. (1) Resting state \ac{FC} matrix $\mathbf{C}^{\text{rest}}$ is used. (2) The task general \ac{FC} $\mathbf{C}^{\text{task}}$ was generated by concatenating the time courses across all tasks before the Pearson's correlation. In (3), we use task-specific \ac{FC} in \autoref{eq:node_strength_m} and \autoref{eq:We_strength} for each hyperedge, denoted as ${\mathbf{C}}^{\text{hyper-task}}$. We implement (4) by using resting state \ac{FC} in \autoref{eq:node_strength_m} and \autoref{eq:We_strength} as described in \autoref{sec:multi_hyper_m}, denoted as ${\mathbf{C}}^{\text{hyper-t-r}}$. For (5), we first generate $\bar{\mathbf{C}}^{\text{task}}_{ij}$ by using task-specific \ac{FC} as $\mathbf{C}_{ij}^{e^k}$, and resting state $\mathbf{C}^{\text{rest}}$ to compute $\tilde w(e^k)$ based on \autoref{eq:node_strength_m} and \autoref{eq:We_strength}. We next applied our proposed local thresholding \cite{Wang-2016-Modularity} on resting state \ac{FC} $\mathbf{C}^{\text{rest}}$ to match with the graph density of $\bar{\mathbf{C}}^{\text{task}}$ at 0.2765, which lies within the normal range of thresholding before subnetwork extraction between [0.2, 0.3] \cite{Van-2008-Normalized}. We then estimate $\mathbf{C}^{\text{t-r}}$ using \autoref{eq:multimodal_hyper}. 
We set free parameters $w1, w2, w3$ to one, and the \textit{stat} to median value based on inner cross-validation. For (6), we generated the $\bar{\mathbf{C}}^{\text{task}}_{ij}$ and thresholded $\mathbf{C}^{\text{rest}}$ as the same way as in (5), then the multisource integration is implemented using a multislice approach \cite{Mucha-2010-Community}, denoted as $\mathbf{C}^{\text{t-r-multislice}}$. We set the weighting for multisource integration $\gamma$ or coupling parameter $\omega$ from 0.01 to 1 at an interval of 0.01. In order to perform fair comparison, $\mathbf{C}^{\text{rest}}$ in method (1) and $\mathbf{C}^{\text{task}}$ in method (2) have also been local thresholded at the graph density of 0.2765. Method (1) to (5) used \ac{Ncuts} and (6) used generalized Louvain as the graph partitioning approach. The number of subnetworks was set to seven given that there are seven tasks available to examine if subnetwork assignments can be related to tasks. We note that setting the number of subnetworks is non-trivial as discussed in the previous section that we leave as future work.
All statistical comparisons are based on the Wilcoxon signed rank test with significance declared at an $\alpha$ of 0.05 with Bonferroni correction.
\subsection{Similarity of \ac{FC} between Resting state and Task data} 
We observed a similarity at \ac{DSC} = 0.7845 between resting state \ac{FC} and task general \ac{FC}, which was generated by concatenating the time courses across all different tasks. For seven specific tasks, the corresponding \ac{DSC} between task-specific \ac{FC} and task general \ac{FC} are 0.8971 for emotion processing, 0.8557 gambling, 0.8676 for language, 0.9043 for motor, 0.8594 for relational processing, 0.8307 for social cognition, and 0.8751 for working memory. This high similarities confirms the findings in \cite{Cole-2014-Intrinsic} that a set of small but consistent changes common across tasks suggests the existence of a task-general network architecture distinguishing task states from rest.

When resting state \ac{FC} is compared to task-specific \ac{FC}, the \ac{DSC} are 0.7193 for emotion processing, 0.7689 for gambling, 0.7390 for language, 0.7067 for motor, 0.7533 for relational processing, 0.7659 for social cognition and 0.7118 for working memory, respectively. The variation of similarities between task-specific and resting state \ac{FC} around a relatively high average level further confirms that the brain’s functional network architecture during task is configured primarily by an intrinsic network architecture which can be present during rest, and secondarily by changes in evoked task-general (common across tasks) and task-specific network \cite{Cole-2014-Intrinsic}.

These findings confirms the close relationship between task and rest, and the support for integrating multitask information into resting state based subnetwork extraction. 
\subsection{Modularity \textit{Q} Value}
\label{subsec:hyper_Q}
Modularity $Q$ value has been used to assess a graph partitioning through reflecting the intra- and inter- subnetwork connection structure of a network \cite{Sporns-2016-Modular}. We observe that $Q$ values of group level subnetwork extraction for method (1)-(6) are 0.1401, 0.1282, 0.1624, 0.1711, 0.2290 and 0.1905 when $\gamma$ and $\omega$ were selected at the highest inter-subject reproducibility. 
% Q: (1) 0.1401 (2) 0.1282 (3) 0.1624 (4) 0.1711 (5) 0.2290... (6)0.1752 at 2; 0.1905 at 7;

\begin{figure}
	\centering
    \includegraphics[width=4.8in]{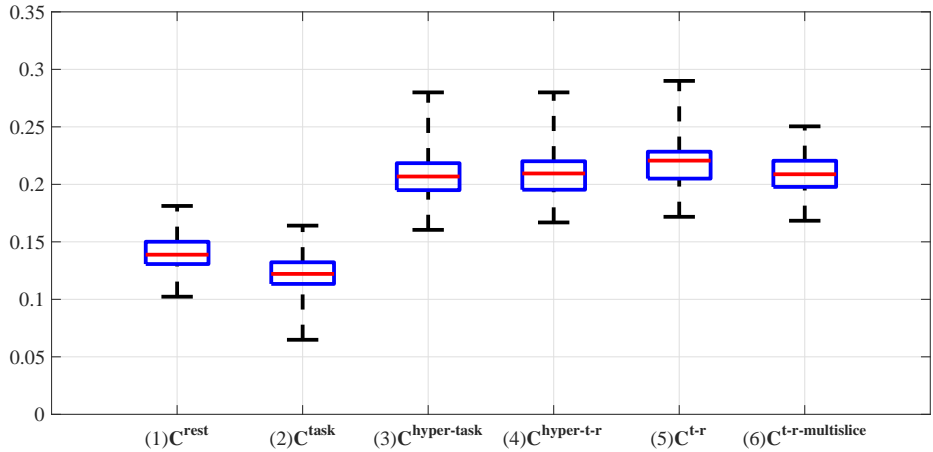}
    \caption{Subject-wise level modularity $Q$ values using Method (1)-(6). For method (5) and (6), parameter $\gamma$ and $\omega$ were selected at the highest inter-subject reproducibility.}
    \label{fig:hyper_Q_subj}
\end{figure}

At the subject-wise level, the modularity $Q$ values estimated from the subnetwork extraction using method (1)-(6) are 0.1397$\pm$0.0142, 0.1234$\pm$0.0159, 0.2072 $\pm$ 0.0199, 0.2094$\pm$0.0189, 0.2183$\pm$0.0192, and 0.2089$\pm$0.0165 respectively, \autoref{fig:hyper_Q_subj}.

We show that the modularity estimated from subnetworks extracted based on simply concatenating task time courses is lower than using resting state data. Using hypergraph framework (3) ${\mathbf{C}}^{\text{hyper-task}}$ and (4) ${\mathbf{C}}^{\text{hyper-t-r}}$ achieves statistically higher modularity values than using either resting state data or simple concatenation of task data. Moreover, incorporating resting state information into the hypergraph framework (5) ${\mathbf{C}}^{\text{t-r}}$ can increase modularity compared to hypergraph method. Multislice integration (6) ${\mathbf{C}}^{\text{t-r-slice}}$ results in a lower modularity than (5) the linear model; however, it still outperforms all the other uni-source methods. Overall, incorporating resting state information explicitly using a weighted combination strategy, \ie method (5) gives a statistically higher modularity than all contrasted methods at $p<10^{-4}$ based on Wilcoxon signed rank test.
We note that the $Q$ values derived here are around 0.2, when the number of the subnetworks was set to seven, \ie the number of tasks. It is relatively low due to the inherent resolution limit of $Q$, \ie $Q$ decreases when the number of subnetworks increases. We explored this direction by achieving the similar level of $Q$  values around 0.3-0.4 when the number of subnetworks decreases to 4 as in \cite{Crossley-2013-Cognitive}.  
\subsection{Inter-subject Reproducibility of Subnetwork Extraction}
\begin{figure}
	\centering
    \includegraphics[width=4.8in]{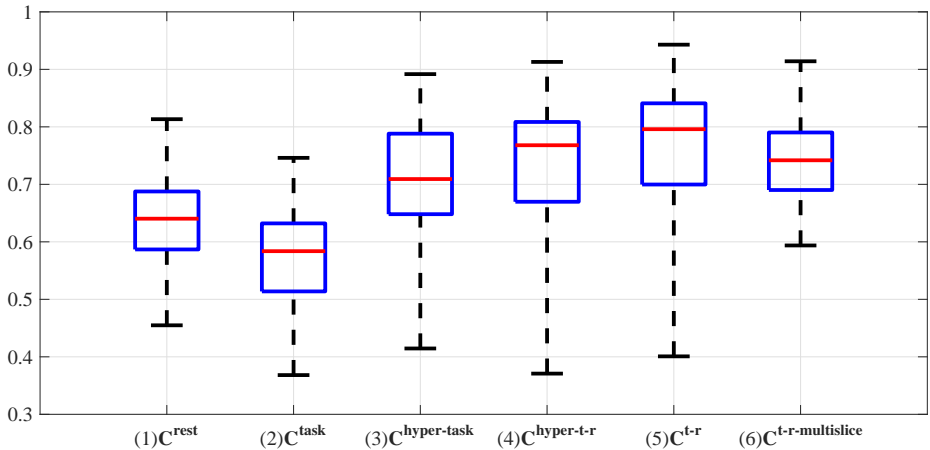}
    \caption{Subject-wise level inter-subject reproducibility of subnetwork extraction using Method (1)-(6). For method (5) and (6), parameter $\gamma$ and $\omega$ were selected at the highest inter-subject reproducibility.}
    \label{fig:hyper_repro_subj}
\end{figure}

We assessed the inter-subject reproducibility by comparing the subnetwork extraction results using subject-wise data against the group level data. The average \ac{DSC} between subject-wise and group level subnetworks across 77 subjects based on methods (1)-(6) are 0.6362$\pm$0.0828, 0.5704$\pm$0.0872, 0.7083$\pm$0.1094, 0.7258$\pm$0.1201, 0.7561$\pm$0.1199, and 0.7406$\pm$0.0725, \autoref{fig:hyper_repro_subj}. We noticed that the reproducibility using resting state \ac{FC} ${\mathbf{C}}^{\text{rest}}$ is higher than simple concatenation of task time courses data ${\mathbf{C}}^{\text{task}}$. It could be that there exist great differences in reaction to stimuli from different subjects, and simple concatenation is hard to discover the higher order relationship between canonical network components. On the other hand, analyzing multi-task information using hypergraph (3) ${\mathbf{C}}^{\text{hyper-task}}$ achieved much higher stability in subnetwork extraction, and incorporating resting information implicitly within the hypergraph (4) ${\mathbf{C}}^{\text{hyper-t-r}}$, or explicit weighted combination (5) ${\mathbf{C}}^{\text{t-r}}$ can even further enhance reproducibility. We note that the weighted combination outperforms multislice integration (6) ${\mathbf{C}}^{\text{t-r-slice}}$, which is still better than all the other uni-source methods. The reason could be that a simple linear model suffices the fusion of task and rest data. Overall, the inter-subject reproducibility derived by (5) ${\mathbf{C}}^{\text{t-r}}$ is statistically higher than all contrasted methods at $p<10^{-4}$ based on Wilcoxon signed rank test.
\subsection{Biological Meaning}
\begin{figure}
	\centering
	\subfloat[Method (1) ${\mathbf{C}}^{\text{rest}}$]{
	    \includegraphics[width=2.5in]{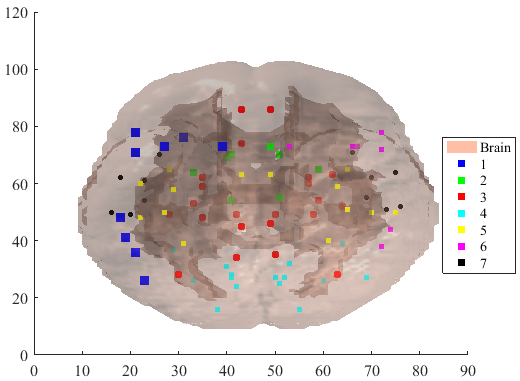}
	}
	\subfloat[Method (2) ${\mathbf{C}}^{\text{task}}$]{
	    \includegraphics[width=2.5in]{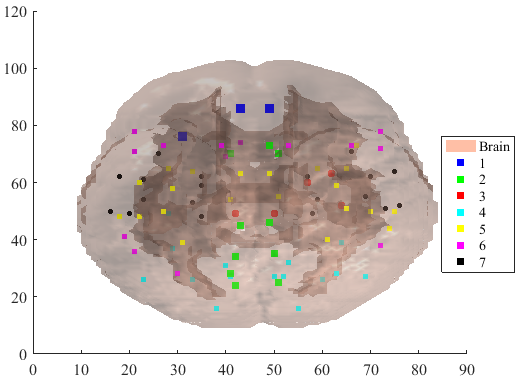}
	}\\
	\subfloat[Method (3) ${\mathbf{C}}^{\text{hyper-task}}$]{
	    \includegraphics[width=2.5in]{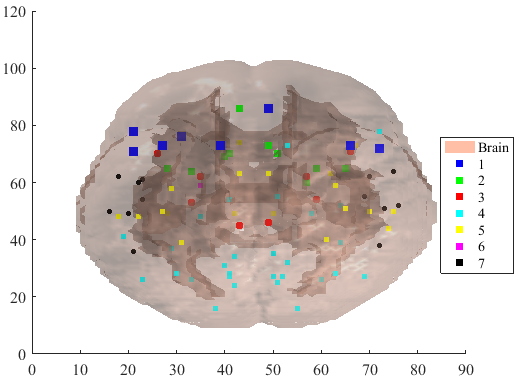}
	}
	\subfloat[Method (4) ${\mathbf{C}}^{\text{hyper-t-r}}$]{
	    \includegraphics[width=2.5in]{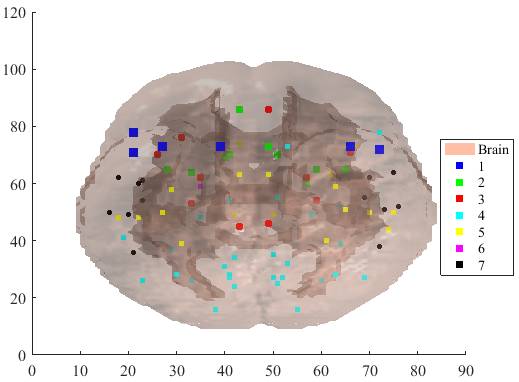}
	}\\
	\subfloat[Method (5) ${\mathbf{C}}^{\text{t-r}}$]{
	    \includegraphics[width=2.5in]{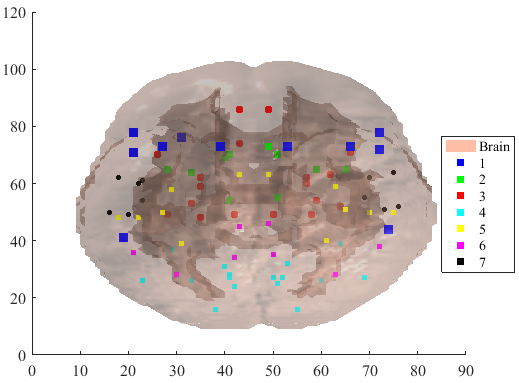}
	}
	\subfloat[Method (6) ${\mathbf{C}}^{\text{t-r-slice}}$]{
	    \includegraphics[width=2.5in]{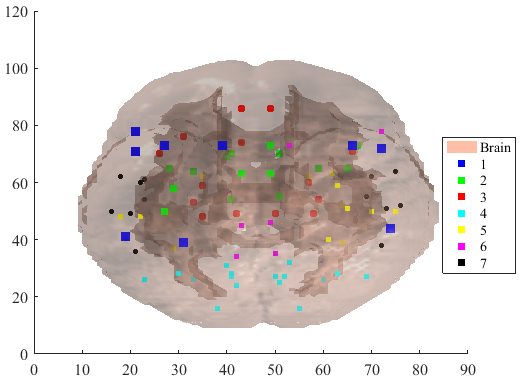}
	}
    \caption{Visualization of subnetworks extraction using methods (1)-(6). The mass center of each \ac{ROI} is plotted in the \ac{MNI} space and colorcoded by the membership of seven subnetworks.}
    \label{fig:hyper_bio_group}
\end{figure}

We next examined the biological meaning of the subnetworks extracted from method (1) - (6), where $\gamma$ was set to 0.5 to report the results when resting state and hypergraph based multitask information are equally combined as an example. Seven subnetworks were extracted based on the number of tasks available. Method (1) detects most of the traditional resting state subnetworks with several false positive and negative detection. The results of method (2) oftentimes combined some important regions from different subnetworks, which lacks biological justifications. Method (3) and (4) generate similar results and both improve the results of method (2) greatly when bringing task dynamics into the subnetwork extraction. Overall, method (5) detects brain regions, which are more biologically meaningful, by combining the intrinsic network architecture from resting state data and the task dynamics based on high-order hypergraph. We report our findings in details as the following and the visualization of subnetwork extraction results can be found in \autoref{fig:hyper_bio_group}. 

Using method (1) based on resting state \ac{FC} alone, subnetwork 1 and 6 are detected as left and right side of a combination of \ac{ECN} and frontoparietal network, which include superior frontal gyrus, middle frontal gyrus, inferior frontal gyrus, posterior supramarginal gyrus, angular gyrus, frontal orbital cortex, and frontal operculum cortex. Method (1) mistakenly classified left inferior lateral occipital cortex and left anterior supramarginal gyrus into the \ac{LECN}. Anterior supramarginal gyrus is part of the somatosensory association cortex, which interprets tactile sensory data and is involved in perception of space and limbs location or language processing, thus it should be included in \ac{DMN} instead of \ac{ECN} \cite{Heinonen-2016-Default}.
On the other hand, our proposed method (5) detects both the left and right sides of most of the anterior portion of \ac{ECN} and posterior supramarginal gyri for subnetwork 1. Using method (5), the left inferior lateral occipital cortex was not include in \ac{ECN}, which is more accurate. Besides, method (5) clustered anterior supramarginal gyrus symmetrically into subnetwork 6, which includes both sides of \ac{PCC}, precuneus, and angular gyrus, comprising most of the posterior portion of \ac{DMN} defined in \cite{Heinonen-2016-Default}. 
As for method (2), the simple concatenation of multitask time courses, subnetwork 1 consists of frontal medial cortex and only the left side of frontal orbital cortex, and subnetwork 6 consists of most of the anterior portion of \ac{ECN}, angular gyrus and only the left posterior supramarginal gyrus, which should be symmetrically included in \ac{DMN}. Besides, there are two other \ac{ROI}s, left subcallosal cortex and left caudate, included in subnetwork 6, which lacks biological meaning. Subnetwork 1 derived from method (3) and (4) both consist of most of the anterior portion of \ac{ECN}, except that method (3) has two more one-sided frontal areas, which makes (4) more biological meaningful (with symmetric results). Subnetwork 6 of method (3) and (4) both consist of one isolate area: left anterior parahippocampal gyrus, which further indicates that there is need to incorporate resting state information into the multitask based on hypergraph framework.

Subnetwork 2 of method (1) includes both sides of \ac{ACC}, caudate, thalamus, putamen and accumbens. Method (5) includes all the same brain regions as method (1) plus one other region, the insula. This subnetwork should be related to the gambling task and emotional processing, which expect to activate \ac{ACC} \cite{Charpentier-2015-Emotion,Koehler-2013-Increased}, ventral striatum (such as thalamus \cite{Koehler-2013-Increased} and accumbens \cite{Limbrick-2017-Neural}), and insula \cite{Leong-2016-White}.
Usually insula is part of the salience network and has been found to play key roles in emotional processing \cite{Cauda-2011-Functional}. 
However, using method (1), the insula was clustered into subnetwork 5 (mostly motor system). Method (2) included right \ac{ACC} and both sides of \ac{PCC}, precuneous, left side of supracalcarine cortex, and accumbens inside subnetwork 2, which seems like a mixture of part of \ac{DMN}, one-sided region from motor system, and one region from gambling system. As for method (3) and (4), they both extracted similar regions for subnetwork 2 as using method (5), except that they missed thalamus and falsely included left frontal medial cortex.
% , which includes Insula, ACC, Putamen, Accumbens, left Frontal Medial Cortex, and right Anterior Parahippocampal Gyrus. 

Subnetwork 3 derived from method (1) includes superior lateral occipital cortex, frontal medial cortex, left subcallosal cortex, \ac{PCC}, precuneous, parahippocampal gyrus, temporal fusiform cortex, brain stem, hippocampus and amygdala. This assignment does not make too much sense by clustering regions from visual, auditory, emotion circuit and frontal system together. Meanwhile, the results using method (5) consists mostly of emotion circuit and social processing, which includes brain stem \cite{Venkatraman-2017-Brainstem}, hippocampus and parahippocampal gyrus \cite{Ohmura-2010-Serotonergic}, amygdala \cite{Zald-2003-Human}, and subcallosal cortex \cite{Laxton-2013-Neuronal}. Method (5) also detected regions related to auditory functions such as temporal pole, which is reasonable since the negative emotion was induced by listening to stories. Subnetwork 3 detected by method (2) includes right anterior parahippocampal gyrus, temporal fusiform cortex and brain stem, which still lacks important brain regions in the emotion circuit. Method (3) detects more biologically meaningful regions than (2), such as hippocampus and amygdala. Using method (4) can even detect more related regions than method (3), such as frontal orbital cortex \cite{Levens-2011-Role}.

Method (1) and (5) detected almost the same brain regions for subnetwork 4, which is the visual system, except that method (5) detected one more region of the inferior lateral occipital cortex, making the results more symmetric. This subnetwork includes inferior lateral occipital cortex, intracalcarine cortex, cuneal cortex, lingual gyrus, occipital fusiform gyrus, temporal occipital fusiform cortex, occipital pole, and supracalcarine cortex. Method (2) detected most of the visual regions except for cuneal cortex and the right supracalcarine cortex. Method (3) and (4) detected extra regions in right \ac{ECN} and auditory system besides all the regions found using (5) in the visual system.

Subnetwork 5 derived from method (1) comprises of the motor system, including precentral gyrus, postcentral gyrus, only the right side of anterior supramarginal gyrus, juxtapositional lobule cortex; and the frontoparietal network including left central opercular cortex, superior parietal lobule, and parietal operculum cortex. Method (5) generated similar results as method (1), only that the results are more symmetric, which include both sides of anterior supramarginal gyrus (part of somatosensory association cortex); and more accurate in terms of frontoparietal network, which includes frontal operculum cortex instead of central opercular cortex. Both method (3) and (4) generated similar regions for subnetwork 5 as well, which includes motor system and frontoparietal network, except that they both included brain stem into this subnetwork. However, method (2) mis-classified insula, putamen and thalamus into the motor and frontal parietal networks. We note that the motor system and frontoparietal network are clustered together, it could be that the working memory tasks recruited both the motor system and frontoparietal network.

As for the subnetwork 7, both method (1) and (5) detected brain regions corresponding to language task and related auditory regions, such as anterior superior temporal gyrus, planum temporale, planum polare, and Heschls gyrus (includes H1 and H2) \cite{Noesselt-2003-Top}. Different from method (1), method (5) included central opercular cortex, which can be explained by how fronto-opercular is related to language \cite{Meyer-2003-Functional}. Method (2) detected some false positive brain regions in the language system such as parahippocampal gyrus, hippocampus and amygdala. Method (3) and (4) correctly clustered all the brain regions into the language network as method (5).
% except for one false positive region of Angular Gyrus.

Method (6) generated similar results compared to method (5), only a couple regions in subnetworks 2 and 5 were switched, a couple regions in subnetwork 6 and 7 were switched, and a couple regions in 1 and 6 were switched. Overall, The subnetwork results derived by method (5) ${\mathbf{C}}^{\text{t-r}}$ have more biological meaning than contrasted methods.

\section{Discussion}
\subsection{Hypergraph encodes higher order nodal relationship}
Subnetwork results derived from methods based on hypergraph achieved higher modularity, higher inter-subject reproducibility, and more reasonable biological meaning than traditional connectivity analysis of pairwise correlation between nodes. These results indicate that hypergraph, which is a natural presentation of multitask activation, can be explored to study higher order relations among the network nodes. The proposed strength informed version of automatic weight setting of the hyperedge incorporates connectivity information to reveal more accurate higher order relationship among nodes rather than just using binary information. 
% We explored the direction of setting the hyperedge weights from the estimation of the possibility of appearance of nodes associated to some important biological roles. 
% Due to the lack of knowledge of the higher order nodal relationship, it is still hard to make straightforward explanation of the higher order relationship between nodes within hyperedges, which requires more related studies from both advanced graph theory and neuroscience. 

\subsection{Multisource Integration Improves Subnetwork Extraction}
We have proved that multisource integration of task and rest information can improve subnetwork extraction compared to using a single source in terms of graphical metrics, inter-subject reproducibility, along with biologically meaningful subnetwork assignments. We note that the implicit integration of rest information into multitask hypergraph achieved less improvements as the explicit integration based on the linear combination. The reason could be that the limited number of tasks available restricts the comprehensive representation of the brain using the hypergraph. Thus, by integrating rest data to compensate possible missing information resulted in overall better outcomes. Another observation is that the linear combination outperforms the multislice community detection, which still performs better than uni-source approaches. Our assumption is that rest and task \ac{FC} are both derived from a single functional modality, which complements each other by revealing the two sides of \ac{FC}, \ie the resting intrinsic side and the activated evoked side. Thus, a simple linear weighted combination would suffice this situation, which outperforms other alternative combination approach in practice. 

\subsection{Limitations and Future Directions}
There are several limitations in our present work. First, our study investigated only seven available tasks with high quality data and decent amount of data per task. This sample of seven tasks is not enough. A possible solution is to have access to both task and rest data from previous task studies or co-activation studies, which covers much wider variety of tasks. At the same time, with much more information from a greater amount of task data, we can devise a reliable automatic manner to determine the integration weighting parameter $\gamma$. The underlying rationale is that with more tasks available, we can rely more on the hypergraph based multitask source, hence the higher $\gamma$.

Secondly, we set the number of the subnetworks to be seven, which corresponds to the number of tasks available. The reason is simply to see if we can associate the subnetwork results to different tasks and gain insights from the findings based on task-induced functions. 
% However, there exist overlapping brain regions that are activated during different tasks, such as emotional processing and gambling. 
In the future, a finer scale of subnetwork extraction using multi-scale hierarchical approach would improve the interpretation of the findings. 

% Moreover, we have explored multimodal integration of resting state \ac{FC} and \ac{AC} in the previous section, and we are to investigate if combining \ac{FC} along with task information with \ac{AC} information will further improve the multimodal subnetwork extraction. We have conducted preliminary studies on applying $\mathbf{C}^{\text{t-r}}$ and \ac{AC} as two modalities using our proposed \ac{RW} based approach. However, more systematic validation approaches and related neuroscience study evidences are required for comparing three sources with two sources.
% - unbiased Q
% - determine $\gamma$, based on the number of tasks
% - different graph densities
% - finer subnetworks; nNets 7 -- 11
% -Why not combine AC into resting FC? Task is more like FC, and have that canonical components ideas. AC is a bit too far away from FC, might introduce some complexity. But our future work can combine the three using the multislice, add a third slice of AC.
% - Future, RW to incorporate 3 sources
\section{Conclusion}
We proposed a high order relation informed approach based on hypergraph to combine the information from multi-task data and resting state data to improve subnetwork extraction. 
% Our assumption is that task data can be beneficial for the subnetwork extraction process, since the repeatedly activated nodes involved in diverse tasks might be the canonical network components which comprise pre-existing repertoires of resting state subnetworks \cite{Park-2014-Graph}. Our proposed high order relation informed subnetwork extraction based on a strength information embedded hypergraph, (1) facilitates the multisource integration for subnetwork extraction, (2) utilizes information on relationships and changes between the nodes across different tasks, and (3) enables the study on higher order relations among brain network nodes.
We demonstrated that fusing task activation, task-induced connectivity and resting state functional connectivity based on hypergraphs improves subnetwork extraction compared to employing a single source from either rest or task data in terms of subnetwork modularity measure, inter-subject reproducibility, along with more biologically meaningful subnetwork assignments.
\section{List of Acronyms}

\begin{acronym}
\acro{AAL}{Automated Anatomical Labeling}
\acro{AC}{Anatomical Connectivity}
\acro{ACC}{Anterior Cingulate Cortex}
\acro{AD}{Alzheimer’s disease}
\acro{BOLD}{Blood Oxygenated Level Dependent}
\acro{BP}{Boundary Point}
\acro{CBP}{connectivity based parcellation}
\acro{Cg}{Cingulate Cortex}
\acro{CIS}{Connected Iterative Scan}
\acro{CNN}{Convolutional Neural Networks}
\acro{COREG}{CO-training with REGularization}
\acro{CPM}{Clique Percolation Method}
\acro{CSA}{Constant Solid Angle}
\acro{CSORD}{Coupled Stable Overlapping Replicator Dynamics}
\acro{DAE}{Deep Auto-Encoder}
\acro{DALYs}{Disability-Adjusted Life Years}
\acro{DBN}{Deep Belief Network} 
\acro{DC}{physical distance}
\acro{DMN}{Default Mode Network}
\acro{dMRI}{Diffusion-weighted Magnetic Resonance Imaging}
\acro{DNNs}{Deep Neural Networks}
\acro{DOF}{degree of freedom}
\acro{DSC}{Dice Similarity Coefficient}
\acro{DTI}{Diffusion Tensor Imaging}
\acro{EC}{Effective Connectivity}
\acro{ECN}{Executive Control Network}
\acro{EEG}{Electroencephalography} 
\acro{EPI}{Echo-Planar Imaging}
\acro{FA}{Fractional Anisotropy}
\acro{FC}{Functional Connectivity}
\acro{fMRI}{Functional Magnetic Resonance Imaging}
\acro{GBD}{Global Burden of Disease}
\acro{GLM}{General Linear Model}
\acro{GMM}{Gaussian Mixture Model}
\acro{GROUSE}{Grassmannian Rank-One Update Subspace Estimation} 
\acro{GT}{Global Thresholding}
\acro{HARDI}{High Angular Resolution Diffusion Imaging}
\acro{HCP}{Human Connectome Project}
\acro{HO}{Harvard-Oxford}
\acro{ICA}{Independent Component Analysis}
\acro{ICC}{Intra-Class Correlation}
\acro{ICNs}{Intrinsic Connectivity Networks}
\acro{IP}{Interior Point}
\acro{IPL}{Inferior Parietal Lobule}
\acro{IQ}{Intelligence Quotient}
\acro{LECN}{Left Executive Control Network}
\acro{LMaFit}{Low-Rank Matrix Fitting}
\acro{LT}{Local Thresholding}
\acro{MC}{Multimodal Connectivity} 
\acro{MCNF}{Matrix Completion with Nonnegative Factorization}
\acro{MCSE}{Multisource Clique-based Subnetwork Extraction}
\acro{MEG}{Magnetoencephalography}
\acro{MITK}{Medical Imaging Interaction Toolkit}
\acro{mmRW}{multi-modal Random Walker}
\acro{MNI}{Montreal Neurological Institute}
\acro{MRI}{Magnetic Resonance Imaging}
\acro{MST-KNN}{minimal spanning tree and k-nearest neighbors}
\acro{MVSC}{MultiView Spectral Clustering}
\acro{Ncuts}{Normalized cuts}
\acro{NMD}{Neighborhood-information-embedded Multiple Density}
\acro{NMF}{Non-negative Matrix Factorization}
\acro{NMI}{Normalized Mutual Information}
\acro{NP}{Non-deterministic Polynomial-time}
\acro{NRMSE}{normalized root-mean-squared-error}
\acro{ODF}{Orientation Distribution Function}
\acro{OSLOM}{Order Statistics Local Optimization Method} 
\acro{PCA}{Principal Component Analysis}
\acro{PCC}{Posterior Cingulate Cortex}
\acro{PD}{Parkinson’s disease}
\acro{PDD}{Principal Diffusion Direction}
\acro{PDF}{Probability Density Function}
\acro{PET}{Positron Emission Tomography}
\acro{RBM}{Restricted Boltzmann Machines}
\acro{RD}{Replicator Dynamics}
\acro{RM}{random parcellation}
\acro{ROI}{region of interest}
\acro{RQ}{Research Questions}
\acro{rs-fcMRI}{Resting State Functional Connectivity based on MRI}
\acro{RW}{Random Walker}
\acro{SAE}{Stacked Auto-Encoder}
\acro{sMRI}{Structural Magnetic Resonance Imaging}
\acro{SNR}{signal-to-noise ratio}
\acro{SORD}{Stable Overlapping Replicator Dynamics}
\acro{SVR}{Support Vector Regression}
\acro{t-fcMRI}{Task Functional Connectivity based on MRI}
\acro{TR}{repetition time}
\acro{WHO}{World Health Organization}
\acro{3D}{three-dimensional}
\end{acronym}

% You can also use \newacro{}{} to only define acronyms
% but without explictly creating a glossary
% 
% \newacro{ANOVA}[ANOVA]{Analysis of Variance\acroextra{, a set of
%   statistical techniques to identify sources of variability between groups.}}
% \newacro{API}[API]{application programming interface}
% \newacro{GOMS}[GOMS]{Goals, Operators, Methods, and Selection\acroextra{,
%   a framework for usability analysis.}}
% \newacro{TLX}[TLX]{Task Load Index\acroextra{, an instrument for gauging
%   the subjective mental workload experienced by a human in performing
%   a task.}}
% \newacro{UI}[UI]{user interface}
% \newacro{UML}[UML]{Unified Modelling Language}
% \newacro{W3C}[W3C]{World Wide Web Consortium}
% \newacro{XML}[XML]{Extensible Markup Language}

\bibliographystyle{splncs}
\bibliography{biblio}
\end{document}